# Supergravity in 10 + 2 Dimensions as Consistent Background for Superstring [1]

Hitoshi  NISHINO

*Department of Physics*
*University of Maryland*
*College Park, 20742-4111, USA*
E-Mail: nishino@umdhep.umd.edu

**Abstract**

We present a consistent theory of $N = 1$ supergravity in twelve-dimensions with the signature $(10, 2)$. Even though the formulation uses two null vectors violating the manifest Lorentz covariance, all the superspace Bianchi identities are satisfied. After a simple dimensional reduction to ten-dimensions, this theory reproduces the $N = 1$ supergravity in ten-dimensions, supporting the consistency of the system. We also show that our supergravity can be the consistent backgrounds for heterotic or type-I superstring in Green-Schwarz formulation, by confirming the fermionic $\kappa$-invariance of the total action. This theory is supposed to be the purely $N = 1$ supergravity sector for the field theory limit of the recently predicted F-theory in twelve-dimensions.

---

[1]This work is supported in part by NSF grant # PHY-93-41926.

# 1. Introduction

There has been some evidence [1] that type IIB and type I or heterotic string theories with no direct M-theory unification in 10+1 dimensions (11D) [2], may arise from a unifying theory called F-theory in $10+2$ dimensions (12D). Like M-theory, F-theory is supposed to provide unified framework for understanding the vacuum structures of those string theories not unified by M-theory [3]. Motivated by this observation, we have established in our previous paper [4] the first superspace formulation for supersymmetric Yang-Mills theory in 12D with the signature (10,2). In this formulation we have presented the set of constraints consistent with all the Bianchi identities (BIs) in superspace, together with the component results, performing also dimensional reductions into the conventional 10D and 4D supersymmetric Yang-Mill theories. More recently there has been super $(2+2)$-brane action formulated in *flat* superspace [5], and purely bosonic sector of F-theory has been also proposed [6]. However, the most important *curved* supergravity background in 12D consistently coupled to the above-mentioned string theories has been still lacking.

In this paper we take the first significant step toward the formulation of supergravity in 12D with two time coordinates. We will experience for the first time in this paper the superspace formulation of supergravity in the presence of null vectors in 12D that violates the manifest Lorentz covariance. The existence of such supergravity theory had been a dream for a long time suggested in different contexts in the past because of small size of Majorana-Weyl spinors in such high dimensions as 12D [7], indicating the possibility of a boson-fermion matching, or as possible $SO(10,2)$ covariant supergravity[2] in the context of recent development of S-theory [8], or higher-dimensional theories with two time coordinates [9]. In this paper we will finally realize the dream in an explicit but amazingly simple way, with the field representations for the algebra of $N=1$ local supersymmetry in 12D. We will see how the system is avoiding possible inconsistency with broken Lorentz symmetry, while keeping some components in the extra dimensions non-vanishing, making the theory non-trivial. Our formulation is also similar to the globally supersymmetric Yang-Mills theory in 12D [4].

As important supporting evidence of the validity of our formulation, we will show how a simple dimensional reduction into $9+1$ dimensions (10D) works, and a set of constraints for $N=1$ supergravity [10] is reproduced. We will also establish a Green-Schwarz superstring sigma-model action [11] with fermionic $\kappa$-symmetry [12], that can couple to our set of superspace backgrounds, with a peculiar constraint lagrangian. We will see that our total action has an additional fermionic symmetry that eliminate half of the freedom of the original fermionic coordinates in 12D superspace.

---

[2]Our formulation is not quite $SO(10,2)$ covariant, as will be shown explicitly.

## 2. Preliminaries with Null Vectors

We introduce two constant null vectors [1][4][5] in our 10+2 dimensions with the signature $(-,+,+,\cdots,+,-)$, defined by

$$(n^a) = (0,0,\cdots,0,+\tfrac{1}{\sqrt{2}},-\tfrac{1}{\sqrt{2}})\ ,\quad (n_a) = (0,0,\cdots,0,+\tfrac{1}{\sqrt{2}},+\tfrac{1}{\sqrt{2}})\ ,$$
$$(m^a) = (0,0,\cdots,0,+\tfrac{1}{\sqrt{2}},+\tfrac{1}{\sqrt{2}})\ ,\quad (m_a) = (0,0,\cdots,0,+\tfrac{1}{\sqrt{2}},-\tfrac{1}{\sqrt{2}})\ . \tag{2.1}$$

with the local Lorentz indices $a,b,\cdots = {}_{(0),(1),\cdots,(9),(11),(12)}$. These null vectors live only in the extra two dimensions for the coordinates $_{(11),(12)}$ with the signature $(+,-)$.

Relevantly, it is convenient to define $\pm$ components associated with the extra dimensions for an arbitrary vector $V_a$:

$$V_\pm \equiv \tfrac{1}{\sqrt{2}}\left(V_{(11)} \pm V_{(12)}\right)\ . \tag{2.2}$$

Now it is clear that the factor $\sqrt{2}$ in (2.1) is to make the identifications such as

$$n_+ = n^- = m_- = m^+ = +1\ ,\quad n_- = n^+ = m_+ = m^- = 0\ , \tag{2.3}$$

valid without extra $\sqrt{2}$ factor.[3] Accordingly we have

$$n^a n_a = m^a m_a = 0\ ,\quad m^a n_a = m^+ n_+ = m_- n^- = +1\ . \tag{2.4}$$

As we introduce the Dirac matrices satisfying $\{\sigma_a,\sigma_b\} = +2\eta_{ab} = 2\,\text{diag.}(-,+,\cdots,+,-)$, we get related important relations such as $\slashed{n}\slashed{m} + \slashed{m}\slashed{n} = +2I$, $\slashed{n}^2 = \slashed{m}^2 = 0$, etc. for the combinations $\slashed{n} \equiv \sigma^a n_a$, $\slashed{m} \equiv \sigma^a m_a$. It is then useful to define the projection operators

$$P_\uparrow \equiv \tfrac{1}{2}\slashed{n}\slashed{m}\ ,\quad P_\downarrow \equiv \tfrac{1}{2}\slashed{m}\slashed{n}\ , \tag{2.5a}$$

$$P_\uparrow P_\uparrow = +P_\uparrow\ ,\quad P_\downarrow P_\downarrow = +P_\downarrow\ ,\quad P_\uparrow + P_\downarrow = +I\ . \tag{2.5b}$$

Depending on purposes, we sometimes switch the indices for the extra dimensions from $a, b, \cdots$ to $+, -$. A useful combination frequent in our formulation is

$$P_{\uparrow\downarrow} \equiv P_\uparrow - P_\downarrow = \sigma^{+-}\ . \tag{2.6}$$

It is also important to note the symmetry

$$(\slashed{n})_{\alpha\dot\beta} = -(\slashed{n})_{\dot\beta\alpha}\ ,\quad (\slashed{m})_{\alpha\dot\beta} = -(\slashed{m})_{\dot\beta\alpha}\ ,$$
$$(P_\uparrow)_{\alpha\beta} = -(P_\downarrow)_{\beta\alpha}\ ,\quad (P_{\uparrow\downarrow})_{\alpha\beta} = +(P_{\uparrow\downarrow})_{\beta\alpha}\ , \tag{2.7}$$

with the Majorana-Weyl spinor indices $\alpha,\beta,\cdots = 1,\cdots,32$ and $\dot\alpha,\dot\beta,\cdots = \dot1,\cdots,\dot{32}$ in 12D. There are other important resulting identities, such as $\slashed{n}P_\uparrow \equiv P_\uparrow\slashed{n} \equiv 0$, $\slashed{m}P_\downarrow \equiv P_\downarrow\slashed{m} \equiv 0$. Unlike $3+1$ dimensions, the dotted and undotted spinorial indices are *not* complex conjugate to each other in 12D [7][4].

---

[3]If we have only integral components in (2.1), the price we have to pay is the involvement of $\sqrt{2}$ in (2.3).

## 3. Superspace Constraints

With these preparations at hand, we are ready to setup the constraints in superspace [13]. Our constraints are similar to those for supersymmetric Yang-Mills theory [4] in the sense that the null vectors are involved explicitly. Our superspace supercurvatures $T_{AB}{}^C$, $R_{ABc}{}^d$ and $G_{ABC}$ satisfy the superspace BIs[4]

$$\nabla_{[A} T_{BC)}{}^D - T_{[AB|}{}^E T_{E|C)}{}^D - \tfrac{1}{2} R_{[AB|e}{}^f (\widetilde{\mathcal{M}}_f{}^e)_{|C)}{}^D \equiv 0 ~, \tag{3.1}$$

$$\tfrac{1}{6} \nabla_{[A} G_{BCD)} - \tfrac{1}{4} T_{[AB|}{}^E G_{E|CD)} \equiv 0 ~. \tag{3.2}$$

These are called $(ABC,D)$ and $(ABCD)$-type BIs. The superspace local Lorentz indices are either bosonic[5] or fermionic with dotted (positive chirality), or undotted (negative chirality) spinorial indices:[6] $A$, $B$, $\cdots = (a,\alpha,\dot{\alpha})$, $(b,\beta,\dot{\beta})$, $\cdots$. However, due to the chiral nature of our chiral superspace, we do *not* have spinorial derivatives with *positive* chirality $\overline{\nabla}_{\dot\alpha}$. Accordingly, the fermionic indices in (3.1) and (3.2) run only the negative chirality $\alpha$, $\beta$, $\cdots$. Introducing engineering dimensions, such as $d=0$ for the $\sigma^a$-matrices, or $d=1/2$ and $d=0$ respectively for fermionic and bosonic fundamental physical fields, *etc.*, we can assign dimensionalities for each BI, *e.g.*, $(\alpha\beta\gamma d)$ and $(\alpha\beta\gamma,d)$-type BIs are at $d=1/2$, while $(\alpha\beta cd)$ and $(\alpha\beta c,d)$ are at $d=1$, *etc.* Our null vectors satisfy the desirable constancy conditions: $\nabla_A n_b \equiv \nabla_A m_b \equiv 0$, as will be proven explicitly later.

The only difference of (3.1) from the usual BIs is the usage of the *modified* Lorentz generator $\widetilde{\mathcal{M}}$ satisfying

$$(\widetilde{\mathcal{M}}_{ab})^{cd} \equiv +\widetilde{\delta}_{[a}{}^c \widetilde{\delta}_{b]}{}^d \quad (\widetilde{\delta}_-{}^a = 0, \text{ otherwise } \widetilde{\delta}_a{}^b = \delta_a{}^b) ~, \tag{3.3a}$$

$$(\widetilde{\mathcal{M}}_{ab})_\alpha{}^\beta \equiv +\tfrac{1}{2} (\sigma_{ab} P_\uparrow)_\alpha{}^\beta ~, \quad (\widetilde{\mathcal{M}}_{ab})_{\dot\alpha}{}^{\dot\beta} \equiv +\tfrac{1}{2} (P_\downarrow \sigma_{ab})_{\dot\alpha}{}^{\dot\beta} ~, \tag{3.3b}$$

where the first one coincides with the usual Lorentz generator, while the second line has extra $P_\uparrow$ and $P_\downarrow$. We will clarify why this structure is needed to satisfy the BIs shortly. It is straightforward to confirm that the generators $\widetilde{\mathcal{M}}_{ij}$ with the 10D indices $i$, $j$, $\cdots = (0), (1), \cdots, (9)$ form the $SO(9,1)$ Lorentz algebra: $[\widetilde{\mathcal{M}}_{ij}, \widetilde{\mathcal{M}}^{kl}] = -\delta_{[i}{}^{[k} \widetilde{\mathcal{M}}_{j]}{}^{l]}$, while $\widetilde{\mathcal{M}}_{-i}$ vanishes identically. Even though the commutators between $\widetilde{\mathcal{M}}_{+-}$ and $\widetilde{\mathcal{M}}_{+i}$ does not obey the $SO(10,2)$ Lorentz algebra, this poses no problem, as will be explained after eq. (3.10). It is also straightforward to confirm that (3.3b) satisfies the Jacobi identity.

Our field content in 12D is the zwölfbein $e_m{}^a$, the gravitino $\psi_m{}^\alpha$, the dilatino $\overline{\chi}_{\dot\alpha}$, antisymmetric tensor $B_{mn}$ and the dilaton $\Phi$. Especially, the gravitino (or the dilatino) has the negative (or positive) chirality: $\sigma_{13}(\psi_m, \overline{\chi}) = (-\psi_m, +\overline{\chi})$. The absence of the *dotted* gravitino $\overline{\psi}_m{}^{\dot{\alpha}}$ in our $N=1$ system is related to the absence of the dotted indices $\dot{\alpha}, \dot{\beta}, \cdots$ in (3.1) and (3.2).

---

[4] To be more precise, there is an additional set of BIs (3.11) to be discussed shortly.

[5] For *curved* bosonic coordinates we use $m$, $n$, $\cdots = 0, 1, \cdots, 9, 11, 12$, essentially in the same way as in [13].

[6] This chirality convention is the same as in [4]. We use, however, the *bars* for dotted spinorial superfields in this paper.

Our results for superspace constrains are summarized as:

$$T_{\alpha\beta}{}^c = (\sigma^{cd})_{\alpha\beta} n_d + (\sigma^{de})_{\alpha\beta} n^c n_d m_e = (\sigma^{cd})_{\alpha\beta} n_d + (P_{\uparrow\downarrow})_{\alpha\beta} n^c , \tag{3.4a}$$

$$G_{\alpha\beta c} = T_{\alpha\beta c} , \tag{3.4b}$$

$$T_{\alpha\beta}{}^\gamma = (P_\uparrow)_{(\alpha|}{}^\gamma (\slashed{n}\overline{\chi})_{|\beta)} - (\sigma^{ab})_{\alpha\beta} (P_\downarrow \sigma_a \overline{\chi})^\gamma n_b \tag{3.4c}$$

$$\nabla_\alpha \Phi = (\slashed{n}\overline{\chi})_\alpha , \tag{3.4d}$$

$$\nabla_\alpha \overline{\chi}_{\dot{\beta}} = -\tfrac{1}{24} (\sigma^{cde} P_\uparrow)_{\alpha\dot{\beta}} G_{cde} + \tfrac{1}{2} (\sigma^c P_\uparrow)_{\alpha\dot{\beta}} \nabla_c \Phi - (\slashed{n}\overline{\chi})_\alpha \overline{\chi}_{\dot{\beta}} , \tag{3.4e}$$

$$T_{\alpha b}{}^c = 0 , \quad T_{\alpha b}{}^\gamma = 0 , \quad G_{\alpha bc} = 0 , \tag{3.4f}$$

$$T_{ab}{}^c = -G_{ab}{}^c , \tag{3.4g}$$

$$R_{\alpha\beta cd} = +(\sigma^{ef})_{\alpha\beta} n_e G_{fcd} , \tag{3.4h}$$

$$\nabla_\alpha G_{bcd} = +\tfrac{1}{2} (\slashed{n}\sigma_{[b} T_{cd]})_\alpha = -\nabla_\alpha T_{bcd} , \quad R_{\alpha bcd} = +(\slashed{n}\sigma_{[c} T_{d]b})_\alpha , \tag{3.4i}$$

$$\nabla_\alpha T_{bc}{}^\delta = -\tfrac{1}{4} (\sigma^{de} P_\uparrow)_\alpha{}^\delta R_{bcde} + T_{bc}{}^\delta (\slashed{n}\overline{\chi})_\alpha + (P_\uparrow)_\alpha{}^\delta (\overline{\chi}\slashed{n} T_{bc}) + (\sigma^{de} T_{bc})_\alpha (P_\downarrow \sigma_d \overline{\chi})^\delta n_e . \tag{3.4j}$$

Here the spinorial inner products are the usual ones, e.g., $(\overline{\chi}\slashed{n} T_{bc}) \equiv \overline{\chi}^{\dot{\alpha}} (\slashed{n})_{\dot{\alpha}}{}^\beta T_{ab\beta} = -\overline{\chi}^{\dot{\alpha}} (\slashed{n})_{\dot{\alpha}\dot{\beta}} T_{ab}{}^{\dot{\beta}}$, etc. All other implicit components such as $G_{\alpha\beta\gamma}$ are zero, as usual. More importantly, due to the chiral nature of our superspace, all other chiral components, such as $T_{\dot{\alpha}\dot{\beta}}{}^c$ or $T_{\alpha\dot{\beta}}{}^c$ are vanishing. Note, however, the existence of the component $R_{AB\gamma}{}^{\dot{\delta}}$ defined by $R_{AB\gamma}{}^{\dot{\delta}} \equiv (1/2) R_{ABc}{}^d (\widetilde{\mathcal{M}}_d{}^c)_\gamma{}^{\dot{\delta}}$, needed for the computation of commutators such as $[\nabla_a, \nabla_b] \overline{\chi}_{\dot{\gamma}} = R_{\alpha b \dot{\gamma}}{}^{\dot{\delta}} \overline{\chi}_{\dot{\delta}}$. Note also that the components $R_{AB\gamma}{}^{\dot{\delta}}$ do not enter the BIs (3.1), because (3.1) is based on the identity $[[\nabla_A, \nabla_B\}, \nabla_C\} + (2 \text{ perms.}) \equiv 0$, with no derivative $\nabla_{\dot{\alpha}}$ involved. This feature is also related to the fact that the dotted superfield $\overline{\chi}_{\dot{\alpha}}$ is not defined like $\nabla_{\dot{\alpha}} \Phi$, but there is an operator $\slashed{n}$ multiplied in (3.4d), avoiding $\nabla_{\dot{\alpha}}$.

In our system, there is further a set of extra constraints dictated by

$$T_{AB}{}^c n_c = 0 , \quad G_{ABc} n^c = 0 , \quad T_{aB}{}^C n^a = 0 , \tag{3.5}$$

$$R_{ABc}{}^d n_d = R_{aBc}{}^d n^a = 0 , \tag{3.6}$$

$$n^a \nabla_a \Phi = 0 , \quad n^a \nabla_a \overline{\chi}_{\dot{\beta}} = 0 , \tag{3.7}$$

$$(\slashed{n}\slashed{n})_\alpha{}^{\dot{\beta}} \overline{\chi}_{\dot{\beta}} = 0 , \quad T_{ab}{}^\gamma (\slashed{n}\slashed{n})_\gamma{}^{\dot{\alpha}} = 0 , \tag{3.8}$$

$$\phi_{Ab}{}^c n_c = \phi_{ab}{}^c n^a = 0 . \tag{3.9}$$

These are locally supersymmetric analog of the constraints $F_{ab} n^b = 0$, $\slashed{n}\overline{\lambda} = 0$, $n^a \nabla_a \overline{\lambda} = 0$ imposed in our previous paper [4]. Their role is to delete some (but not all) extra components that are non-physical after dimensional reductions into lower dimensions. By this prescription, we can get the right chiral field content for the resulting 10D supergravity, as will be seen shortly. It is also to be stressed that these extra constraints will not eliminate all the freedom of relevant superfields in the extra dimensions. For example, even though the component $G_{ab-}$ is eliminated by (3.5), the other extra component $G_{ab+}$ is still alive in our theory. This feature is crucial to have non-trivial system in 12D that is distinct from merely a rewriting of 10D system. This is similar to the supersymmetric Yang-Mills case [4].

Compared with our supersymmetric Yang-Mills theory in 12D [4], there is an extra term in (3.4a) which is needed to satisfy the $d = 3/2$ $(\alpha bcd)$, $(\alpha bc, d)$ and $(a\beta\gamma, \delta)$ BIs. Since this term is proportional to $n^c$, it has no effect in the commutator $\{\nabla_\alpha, \nabla_\beta\}$ acting directly on physical superfields, but it has new effect, only when $T_{\alpha\beta}{}^c$ is involved in the terms like $T_{bc}{}^\epsilon T_{\epsilon\alpha}{}^d$ in the $(\alpha bc, d)$ BI mentioned above. It is also important to note that the presence of this term will not affect our previous supersymmetric Yang-Mills system [4].

The structure (3.3a) is crucial for the covariant constancy of the null vectors:

$$\nabla_M n^- = \partial_M n^- + \tfrac{1}{2}\phi_M{}^{ab}(\mathcal{M}_{ba})^{-+}n_+ = 0 \ , \quad \nabla_M m^+ = \partial_M m^+ + \tfrac{1}{2}\phi_M{}^{ab}(\mathcal{M}_{ba})^{+-}m_- = 0 \ . \tag{3.10}$$

Especially, the last terms vanish due to $\widetilde{\delta}_-{}^a = 0$ in (3.3). Note that we are *not* imposing conditions like $\phi_{Ma}{}^b m_b = 0$ directly, that would delete *all* the extra components in $\phi_{Ma}{}^b$, reducing the theory to 10D supergravity. As has been mentioned, the irregular feature of the commutator $[\widetilde{\mathcal{M}}_{+-}, \widetilde{\mathcal{M}}_{+i}]$ poses no inconsistency, because when they are multiplied by the Lorentz connection, no extra component will be effectively left over. In fact, consider $\phi_A{}^{ab}\widetilde{\mathcal{M}}_{ab} = \phi_A{}^{ij}\widetilde{\mathcal{M}}_{ij} + 2\phi_A{}^{+i}\widetilde{\mathcal{M}}_{+i} + 2\phi_A{}^{+-}\widetilde{\mathcal{M}}_{+-} + 2\phi_A{}^{-i}\widetilde{\mathcal{M}}_{-i}$, where the last term vanishes due to $\widetilde{\mathcal{M}}_{-i} = 0$, while the second and third terms also disappear thanks to (3.9). Additionally, due to the extra constraints (3.6), the commutators $[\nabla_A, \nabla_B\}$ acting both on $m_c$ and $n_c$ vanish consistently. This system cleverly avoids inconsistency within the 10D sub-manifold, while maintaining the non-vanishing superfield components such as $\phi_M{}^{-i}$ in the extra directions.

In addition to the BIs in (3.1) and (3.2), there is another set of BIs called $R$-BIs of the type $(ABCd, e)$:

$$\nabla_{[A}R_{BC)d}{}^e - T_{[AB|}{}^E R_{E|C)d}{}^e \equiv 0 \ . \tag{3.11}$$

Usually these BIs are automatically satisfied, once (3.1) holds [14]. In our system, however, this is non-trivial due to our modified Lorentz generators (3.3). Fortunately, we can easily confirm that the $(\alpha\beta\gamma d, e)$ BI at $d = 3/2$ is satisfied, while for the $(\alpha\beta cd, e)$ BI at $d = 2$, we can show that the lemma used in ref. [14] is still valid even with our modified Lorentz generators by explicit computation. The remaining $R$-BIs at $d \geq 5/2$ will not give any non-trivial consistency check. Thus we can conclude that all the BIs in our superspace are satisfied.

There is a technical but important property about $\nabla_A$. Since our Lorentz generators are modified, *not* all the 'constant' matrices like $C_{\alpha\beta}$ are commuting with $\nabla_A$. Because of the non-vanishing commutator $[\widetilde{M}_{ab}, C_{\alpha\beta}]$, the ups/downs of spinorial indices need special care in general. Fortunately, however, we can show that *all* the constant matrices involved in the constraints (3.4) are commuting with $\nabla_A$, *e.g.*, $[\nabla_A, T_{\alpha\beta}{}^c] = 0$. In this connection, important relations are such as

$$[\nabla_A, (\not{n})_\alpha{}^{\dot\beta}] = [\nabla_A, (\not{m})_\alpha{}^{\dot\beta}] = [\nabla_A, (P_\uparrow)_\alpha{}^\beta] = [\nabla_A, (P_\downarrow)_\alpha{}^\beta] = 0 \quad, \qquad (3.12a)$$

$$[\nabla_A, (\sigma_{cd} P_\uparrow)_\alpha{}^\beta] = [\nabla_A, (P_\downarrow \sigma_{cd})_{\dot\alpha}{}^{\dot\beta}] = 0 \quad, \qquad (3.12b)$$

$$[\nabla_A, (\not{n} \sigma^a P_\downarrow)_{\alpha\beta}] = [\nabla_A, (P_\uparrow \sigma^a \not{n})_{\alpha\beta}] = 0 \quad. \qquad (3.12c)$$

Using these relations, we can show that *e.g.*, all the 'constant' matrices in (3.4j) commute with $\nabla_A$, so that we can directly apply $\nabla_A$ to $T_{bc}{}^\delta$ or $\overline\chi_{\dot\beta}$.

Originally, the modification of our Lorentz generators (3.3) was required by the $G$-linear terms in the $(\alpha\beta\gamma,\delta)$ BI at $d=1$. Without this modification, we found that terms like $(\sigma^{ab})_{(\alpha\beta|}(\sigma^{fc})_{|\gamma)}{}^\delta G_{acd} n_f n_b m^d$ are left over. We found that these terms are completely cancelled, only when the Lorentz generator is modified like (3.3).

Even though we skip the details, we can perform various mutual consistency checks of the extra constraints (3.5) - (3.9) with supersymmetry. For example, we can confirm easily that the derivative of (3.8) vanishes: $\nabla_\alpha [\,(\not{n})_\gamma{}^{\dot\beta} T_{bc}{}^\gamma\,] = 0$, *etc.*

We next give our superfield equations from the BIs in (3.1) and (3.2) at $d \geq 3/2$. As usual, at $d = 3/2$, (3.4i) satisfies the $(\alpha bcd)$ and $(\alpha bc, d)$ BIs. From the $(\alpha\beta\gamma, \delta)$ BI we get the gravitino/dilatino superfield equation. In fact, if we let $X_{ABC}{}^D$ be the l.h.s. of the $(ABC, D)$ BI in (3.1), by evaluating the component $X_{\alpha\beta\gamma}{}^\beta$, we get

$$(\sigma^{bc})_{\alpha\beta} T_{ab}{}^\beta n_c - 2(\not{n})_\alpha{}^{\dot\beta} \nabla_a \overline\chi_{\dot\beta} = 0 \quad. \qquad (3.13)$$

At $d = 2$, (3.4j) will satisfy the $(\alpha bc, \delta)$ BI. Using this and taking a spinorial derivative of (3.13) in the combination $(\sigma_{bc})^{\beta\alpha} \nabla_\beta [\,(3.13)_{a\alpha}\,]$, we get the gravitational superfield equation

$$R_{a[b|} n_{|c]} + 4 \nabla_a \nabla_{[b|} \Phi n_{|c]} - 4 (\overline\chi \not{n} T_{a[b|}) n_{|c]} = 0 \quad. \qquad (3.14)$$

Finally, the important relationship

$$R_{[ab]} = -\nabla_c G_{ab}{}^c \quad, \qquad (3.15)$$

comes out of the $(abc, d)$ BI *via* the contraction $X_{abc}{}^c$. All of these are just parallel to the 10D case [10], and are also analogous to the corresponding ones for supersymmetric Yang-Mills multiplet in [4].

Some remarks are in order for these results. There seems to be *a priori* no systematic method to fix the constraint system we obtained in this paper, due to the lack of Lorentz covariance inherent in the system. Our important guiding principle that led us to these expressions is to reproduce the well-known $N = 1$ supergravity in 10D after simple dimensional reduction to be explained shortly. Before reaching our results (3.4) and (3.5), we have tried many different options, such as assigning the same chirality in 12D for $\psi_m{}^\alpha$ and $\chi_\alpha$, modifying the $R$- or $G$-BIs by Chern-Simons forms, or requiring no $G$-BI at all, assuming that the dilaton and the dilatino in 10D would come out of the extra components in the zwölfbein and the gravitino, all in vain. For example, the idea of getting the dilatino out of the 12D gravitino failed, because the $\chi$-transformation rule did not arise right out of the gravitino in the dimensional reduction process we deal with next. All of these trials seem to indicate the dimensional reduction is a key guiding principle to fix the right constraint set in 12D. It is a kind of "oxidation" procedure from 10D to 12D that led us to our results. In particular, fixing the right form for $T_{\alpha\beta}{}^\gamma$ seems to be the crucial step. Relevantly, we found that the modified generator (3.3) for the dotted indices are crucial, in order to get the right gravitino/dilatino superfield equation (3.13) at $d = 3/2$ that reproduces 10D superfield equation (4.8) to be discussed.

Compared with the supersymmetric Yang-Mills case in [4], we noticed that no auxiliary fields, such as the $\chi$-field (in the notation of ref. [4]) are needed for the satisfaction of BIs. We do not know yet the reason why we need no such auxiliary fields for supergravity.

Before ending this section, we give the explicit component field supersymmetry transformation rule obtained from our constraints (3.4) using the method in ref. [13]:

$$
\begin{aligned}
\delta_Q e_m{}^a &= + \left(\epsilon\sigma^{ab}\psi_m\right) n_b + \left(\epsilon P_{\uparrow\downarrow}\psi_m\right) n^a \ , \quad \delta_Q \Phi = -\left(\epsilon\slashed{n}\overline{\chi}\right) \ . \\
\delta_Q \psi_m{}^\alpha &= D_m \epsilon^\alpha + (P_\downarrow\epsilon)^\alpha \left(\overline{\chi}\slashed{n}\psi_m\right) + (P_\downarrow\psi_m)^\alpha \left(\epsilon\slashed{n}\overline{\chi}\right) - (P_\downarrow\sigma_a\overline{\chi})^\alpha \left(\epsilon\sigma^{ab}\psi_m\right) n_b \ , \\
\delta_Q B_{mn} &= + \left(\epsilon\sigma_{[m}{}^b\psi_{n]}\right) n_b - \left(\epsilon P_{\uparrow\downarrow}\psi_{[m}\right) n_{n]} \ , \\
\delta_Q \overline{\chi}_{\dot\alpha} &= + \tfrac{1}{24}(P_\downarrow\sigma^{mnr}\epsilon)_{\dot\alpha} G_{mnr} + \tfrac{1}{2}(P_\downarrow\sigma^m\epsilon)_{\dot\alpha} \nabla_m\Phi - \overline{\chi}_{\dot\alpha}\left(\epsilon\slashed{n}\overline{\chi}\right) \ ,
\end{aligned}
\tag{3.16}
$$

The Lorentz connection $\phi_m{}^{ab}$ involved in $D_m\epsilon^\alpha \equiv \partial_m\epsilon^\alpha + (1/4)\phi_{ma}{}^b (\sigma_b{}^a P_\uparrow\epsilon)^\alpha$ has the torsion $T_{ab}{}^c = -G_{ab}{}^c$ as well as the $\psi$-torsion with our $T_{\alpha\beta}{}^c$ in (3.4a) [13]. Compared with the supersymmetric Yang-Mills case [4], the transformation of $e_m{}^a$ now has the null vector, while the leading term in $\delta_Q \psi_m{}^\alpha$ does *not*.

## 4. Dimensional Reduction into 10D

Next important confirmation is to show that our system reproduces already known results in lower dimensions such as 10D. In this paper we perform a simple dimensional reduction from 12D to 10D, paying special attention to the null vectors we introduced. Other than the role played by the null vectors, our prescription of dimensional reduction in superspace is similar to that in ref. [15].

It has been known that there are innumerably many mutually equivalent sets of constraints for $N = 1$ supergravity in 10D due to the freedom of super Weyl-rescaling. For simplicity of computation, we use what is called "$\beta$-function-favored constraints" (BFFC) [10], which is the simplest set of constraints among possible constraint sets, arranged by appropriate super Weyl-rescalings. By this choice, our constraint system is drastically simplified in 10D, that saves considerable effort in our dimensional reduction.

To distinguish the 10D quantities from the original 12D ones, we introduce the *hat* symbols on the fields and indices in 12D, only within this section. This procedure is a supergravity analog of the similar one we have performed in ref. [4]. We first setup the dimensional reduction rule for the $\sigma$-matrices as in [4], as

$$\widehat{\sigma}_{\hat{a}} = \begin{cases} \widehat{\sigma}_a = \sigma_a \otimes \tau_3 \ , \\ \widehat{\sigma}_{(11)} = I \otimes \tau_1 \ , \\ \widehat{\sigma}_{(12)} = -I \otimes i\tau_2 \ , \end{cases} \tag{4.1}$$

where $\tau_i$ ($i = 1, 2, 3$) are the usual Pauli matrices, and all the *non-hatted* quantities and indices are for 10D. Accordingly, the dimensional reduction for the charge conjugation matrix $\widehat{C}$ and $\widehat{\sigma}_{13}$ in 12D are [4]

$$\widehat{C} = C \otimes \tau_1 \ , \quad \widehat{\sigma}_{13} = \sigma_{11} \otimes \tau_3 \ , \tag{4.2}$$

with the charge conjugation matrix $C$ in 10D. Subsequently, $\widehat{\rlap{/}n}$ and $\widehat{\rlap{/}m}$ satisfy the relationships such as

$$(\widehat{\rlap{/}n})_{\hat{\alpha}}{}^{\hat{\beta}} = (\widehat{\sigma}^+)_{\hat{\alpha}}{}^{\hat{\beta}} = \sqrt{2} I \otimes \begin{pmatrix} 0 & 1 \\ 0 & 0 \end{pmatrix} \ , \quad (\widehat{\rlap{/}m})_{\hat{\alpha}}{}^{\hat{\beta}} = (\widehat{\sigma}^-)_{\hat{\alpha}}{}^{\hat{\beta}} = \sqrt{2} I \otimes \begin{pmatrix} 0 & 0 \\ 1 & 0 \end{pmatrix} \ ,$$

$$\widehat{P_\uparrow} = I \otimes \begin{pmatrix} 1 & 0 \\ 0 & 0 \end{pmatrix} \ , \quad \widehat{P_\downarrow} = I \otimes \begin{pmatrix} 0 & 0 \\ 0 & 1 \end{pmatrix} \ . \tag{4.3}$$

In this representation, the operations of $\widehat{P_\uparrow}$ and $\widehat{P_\downarrow}$ are transparent. We next require that the $\widehat{\chi}$ and the gravitino field strength $\widehat{T}_{ab}$ have the components

$$\left( \widehat{\overline{\chi}}_{\hat{\alpha}} \right) = \begin{pmatrix} \widehat{\overline{\chi}}_{\alpha\uparrow} \\ \widehat{\overline{\chi}}_{\alpha\downarrow} \end{pmatrix} = \begin{pmatrix} 0 \\ \chi_\alpha \end{pmatrix} \ , \quad \left( \widehat{T}_{\hat{a}\hat{b}}{}^{\hat{\gamma}} \right) = \left( \widehat{T}_{\hat{a}\hat{b}}{}^{\gamma\uparrow}, \widehat{T}_{\hat{a}\hat{b}}{}^{\gamma\downarrow} \right) = (T_{\hat{a}\hat{b}}{}^{\gamma}, 0) \ , \tag{4.4}$$

where $\chi_\alpha$ and $T_{ab}{}^\gamma$ are to be the resulting 10D superfields. From now on, we use the index $\uparrow$ (or $\downarrow$) for the first (or second) component of a 12D spinor decomposed into the Pauli matrix space in (4.1). These ansätze are consistent with our extra constraints (3.5) - (3.9), as well as their 12D chiralities, *via* (4.2), which had been fixed in such a way that the resulting 10D supergravity theory has the right chiralities.

We can now easily show that all the constraints (3.4) and the superfield equations (3.13) - (3.15) are reduced into 10D following the simple dimensional reduction in superspace [15], respecting also $\widehat{\nabla}_+ = \widehat{\nabla}_- \equiv 0$, $\widehat{T}_{+B}{}^C = 0$, $\widehat{R}_{+Bc}{}^d = \widehat{R}_{AB+}{}^d = 0$, $\widehat{G}_{+AB} = 0$, as usual.

For example, the dimensional reduction for the $\hat{c} = c, \hat{\alpha} = \alpha\uparrow, \hat{\beta} = \beta\uparrow$-component of $\widehat{T}_{\hat{\alpha}\hat{\beta}}{}^{\hat{c}}$ becomes

$$\widehat{T}_{\alpha\uparrow\beta\uparrow}{}^c = (\widehat{\sigma}^{c+})_{\alpha\uparrow\beta\uparrow} = (\widehat{\sigma}^c \widehat{\rlap{/}n})_{\alpha\uparrow\beta\uparrow} = (\widehat{\sigma}^c)_{\alpha\uparrow}{}^{\hat{\gamma}} (\widehat{\rlap{/}n})_{\hat{\gamma}}{}^{\hat{\delta}} \widehat{C}_{\hat{\delta}\beta\uparrow}$$

$$= \left[ (\sigma^c)_\alpha{}^\gamma \otimes \begin{pmatrix} 1 & 0 \\ 0 & -1 \end{pmatrix} \delta_\gamma{}^\delta \otimes \begin{pmatrix} 0 & \sqrt{2} \\ 0 & 0 \end{pmatrix} C_{\delta\beta} \begin{pmatrix} 0 & 1 \\ 1 & 0 \end{pmatrix} \right]_{\uparrow\uparrow} \tag{4.5}$$

$$= \sqrt{2}(\sigma^c)_{\alpha\beta} \otimes \begin{pmatrix} 1 & 0 \\ 0 & 0 \end{pmatrix}_{\uparrow\uparrow} = \sqrt{2}(\sigma^c)_{\alpha\beta} \equiv T_{\alpha\beta}{}^c \ ,$$

in agreement with the 10D constraint [10], up to a non-essential rescaling factor $\sqrt{2}$. Similarly, if we look into the $\hat{\alpha}=\alpha\uparrow$, $\hat{\beta}=\beta\uparrow$, $\hat{\gamma}=\gamma\uparrow$-component of $\widehat{T}_{\hat{\alpha}\hat{\beta}}{}^{\hat{\gamma}}$:

$$\widehat{T}_{\alpha\uparrow\beta\uparrow}{}^{\gamma\uparrow} \to T_{\alpha\beta}{}^{\gamma} = \sqrt{2}\left[\delta_{(\alpha}{}^{\gamma}\chi_{\beta)} - (\sigma^a)_{\alpha\beta}(\sigma_a\chi)^{\gamma}\right] \quad, \tag{4.6}$$

again in agreement with the 10D result [10]. Parallel procedures work for other constraints in (3.4), reproducing BFFC [10]. We can also make it sure that all the superfield equations in 10D are reproduced by similar procedures from (3.13) and (3.14). In fact, by looking into the $\hat{a}[\hat{b}\hat{c}] = a[b+]$-component of (3.14), we get the 10D gravitational superfield equation in BFFC [10]:

$$R_{ab} + 4\nabla_a\nabla_b\Phi - 4\sqrt{2}(\overline{T}_{ab}\chi) = 0 \quad, \tag{4.7}$$

while all other components such as $\hat{a}[\hat{b}\hat{c}] = a[b-]$ are trivially satisfied. Here we performed the usual identification $\widehat{R}_{ab} = R_{ab}$, $\widehat{\Phi} = \Phi$, *etc.* The $G$-superfield equation is also contained herein by the relationship $R_{[ab]} = -\nabla_c G_{ab}{}^c$ [10]. Similarly from the $\hat{a} = a$, $\hat{\alpha} = \alpha\uparrow$-component of (3.13) we get the gravitino/dilatino superfield equation

$$\sigma^b T_{ab} + \nabla_a\chi = 0 \quad, \tag{4.8}$$

in agreement with [10].

## 5. Green-Schwarz Superstring $\sigma$-Model Action

We have thus far established the superspace formulation of $N=1$ supergravity in 12D. However, we still need to see if such a system can be consistent backgrounds for heterotic or type-I superstring theory. Here we give the action for Green-Schwarz superstring $\sigma$-model, and confirm its $\kappa$-invariance. Due to our extra coordinates, we need a peculiar constraint lagrangian, and we show how such term is consistent with the $\kappa$-symmetry. This confirmation provides supporting evidence for our system to be regarded as the 12D origin of heterotic or type-I superstring theories.

Our total action $I$ is composed of the $\sigma$-model action $I_\sigma$, the Wess-Zumino-Novikov-Witten term $I_B$, and the constraint action $I_\Lambda$:

$$I \equiv I_\sigma + I_B + I_\Lambda \quad, \tag{5.1}$$

$$I_\sigma \equiv \int d^2\sigma \left[V^{-1}\eta_{ab}\Pi_+{}^a\Pi_-{}^b\right] \quad, \tag{5.2}$$

$$I_B \equiv \int d^2\sigma \left[V^{-1}\Pi_+{}^A\Pi_-{}^B B_{BA}\right] \quad, \tag{5.3}$$

$$I_\Lambda \equiv \int d^2\sigma \left[V^{-1}\Lambda_{++}\left(\Pi_-{}^a n_a\right)\left(\Pi_-{}^b m_b\right) + V^{-1}\widetilde{\Lambda}_{++}\left\{(\Pi_-{}^a n_a)^2 + (\Pi_-{}^a m_a)^2\right\}\right] \quad, \tag{5.4}$$

where we use the zweibein $V_+{}^i$, $V_-{}^i$ for the 2D world-sheet with the coordinates $\sigma^i$, and $V \equiv \det(V_{(i)}{}^j)$ with the curved indices $i, j, \cdots$ and the flat light-cone indices $(i), (j), \cdots = +, -$, while $\Pi_{(i)}{}^A \equiv V_{(i)}{}^j \left(\partial_j Z^M\right) E_M{}^A$ with the vielbein $E_A{}^M$ for the 12D superspace with its supercoordinates $Z^M$. The action $I_\Lambda$ has non-propagating Lagrange multipliers $\Lambda_{++}$ and $\widetilde{\Lambda}_{++}$, deliberately chosen such that their field equations get rid of the unwanted contribution to the conformal anomaly, or to the $\kappa$-variation of our total action, as will be seen next.

Our action has two fermionic symmetries with the parameters $\overline{\kappa}_+$ and $\eta$, the former of which is analogous to the 10D case [16][17]:[7]

$$\delta V_+{}^i = \overline{\kappa}_+{}^{\dot{\alpha}}(\sigma^c)_{\dot{\alpha}}{}^\beta \Pi_{+\beta} n_c V_-{}^i \equiv (\overline{\kappa}_+ \not{n} \Pi_+) V_-{}^i \ , \quad (\not{n})_\alpha{}^{\dot{\beta}} \overline{\kappa}_{+\dot{\beta}} \equiv (\not{n}\overline{\kappa}_+)_\alpha = 0 \ , \quad (5.5a)$$

$$\delta V_-{}^i = 0 \ , \quad \delta(V^{-1}) = 0 \ , \quad \delta \overline{E}^{\dot{\alpha}} = \delta E^a = 0 \ , \tag{5.5b}$$

$$\delta E^\alpha = \tfrac{1}{2} (\sigma_a)^{\alpha\dot{\beta}} \overline{\kappa}_{+\dot{\beta}} \Pi_-{}^a + (P_\uparrow)^{\alpha\beta} \eta_\beta \equiv \tfrac{1}{2} (\not{\Pi}_- \overline{\kappa}_+)^\alpha + (P_\uparrow \eta)^\alpha \ , \tag{5.5c}$$

$$\delta \Lambda_{++} = -2 (\overline{\kappa}_+ \not{n} \Pi_+) \ , \quad \delta \widetilde{\Lambda}_{++} = 0 \ , \tag{5.5d}$$

where as usual $\delta E^A \equiv (\delta Z^M) E_M{}^A$. Our parameter $\overline{\kappa}_+$ is subject to the extra constraint in (5.5a), while $\eta$ is unconstrained.

We now confirm the $\kappa$-invariance of the total action. We can easily show that

$$\delta_\kappa (I_\sigma + I_B) = +2 V^{-1} (\overline{\kappa}_+ \not{n} \Pi_+)(\Pi_-{}^a n_a)(\Pi_-{}^b m_b) \ , \tag{5.6}$$

after cancellations of various terms. We have used relations like the 10D case, such as $V \delta_\kappa I_B = -\Pi_-{}^B \Pi_+{}^A (\delta_\kappa E^C) G_{CAB}$, as well as proper 12D relations such as $[P_{\uparrow\downarrow}, \sigma^a] = 2m^a \not{n} - 2n^a \not{m}$. Notice that (5.6) is just proportional to $(\Pi_-{}^a n_a)(\Pi_-{}^a m_a)$. Now for $I_\Lambda$, we use relations like $\delta_\kappa(\Pi_-{}^a n_a) = \delta_\kappa(\Pi_-{}^a m_a) = 0$, $T_{\alpha\beta}{}^c m_c = 0$ to get

$$\delta_\kappa I_\Lambda = -2 V^{-1} (\overline{\kappa}_+ \not{n} \Pi_+)(\Pi_-{}^a n_a)(\Pi_-{}^b m_b) \ . \tag{5.7}$$

By adding (5.7) to (5.6), it is clear that the variation of our total action vanishes:

$$\delta_\kappa I = \delta_\kappa (I_\sigma + I_B + I_\Lambda) = 0 \ . \tag{5.8}$$

In a similar fashion, using $\delta_\eta I_\Lambda = 0$, *etc.*, we can show the $\eta$-invariance

$$\delta_\eta I = 0 \ , \tag{5.9}$$

whose meaning will be explained shortly.

The significance of $I_\Lambda$ is seen as follows. Without $I_\Lambda$, we have to put a constraint $\Pi_-{}^a n_a = 0$ or $\Pi_-{}^a m_a = 0$ by hand for the $\kappa$-invariance. However, since this equation has a first derivative on the 2D world-sheet, such an equation is no longer regarded as a constraint, but is a "field equation" that should *not* be imposed upon the invariance check.

---

[7]There is an alternative $\kappa$-invariance of our action with a parameter $\lambda_+$ whose transformation rule is obtained by the redefinition $\overline{\kappa}_{+\dot{\alpha}} = (\not{n}\lambda_+)_{\dot{\alpha}}$ with the constraint $(\not{n}\lambda_+)_{\dot{\alpha}} = 0$.

We also mention that $I_\Lambda$ or the extra bosonic coordinates in 12D will not contribute to the conformal anomaly in 2D world-sheet. Consider the field equations of $\Lambda_{++}$ and $\widetilde{\Lambda}_{++}$:

$$\left(\Pi_-{}^a n_a\right)\left(\Pi_-{}^b m_b\right) = 0 \ , \quad \left(\Pi_-{}^a n_a\right)^2 + \left(\Pi_-{}^a m_a\right)^2 = 0 \ , \tag{5.10}$$

which are equivalent to

$$\Pi_-{}^a n_a = 0 \ , \quad \Pi_-{}^a m_a = 0 \ . \tag{5.11}$$

Eq. (5.11) physically implies that the extra coordinates $X^\pm$ are independent of the world-sheet coordinate $\sigma^-$. Once $\Pi_-{}^\pm$ vanishes, $\Pi_+{}^\pm$ in $I_\sigma$, $I_B$ or $I_\Lambda$ will not contribute to the energy-momentum, and therefore not to the conformal anomaly. The contribution of $I_\Lambda$ to the $X^m$-field equation also vanishes, due to the factor $\Pi_-{}^\pm$ always involved, while $\Lambda_{++}$ and $\widetilde{\Lambda}_{++}$ themselves disappear from field equations, and therefore will not enter the classical string spectrum. This feature is exactly the same as the constraint actions in ref. [18]. Additionally we can see that all the extra components in the $X^m$-field equations are now satisfied under (5.11).

Finally the $\eta$-invariance with (5.5c) implies the redundancy of half of the total 32 components of the superspace coordinates $\theta^\mu$ in 12D, like a gauge symmetry. After all, thanks to the $\kappa$- and $\eta$-fermionic invariances, the original 32 degrees of freedom of $\theta^\mu$ are reduced like $32 \to 16 \to 8$, in accordance with the 10D Green-Schwarz formulation [16].

## 6. Concluding Remarks

In this paper we have presented a remarkable system of supergravity in 12D for the first time. We have shown how the superspace BIs are satisfied in the presence of null vectors. We have also confirmed that the BIs yield the superfield equations in 12D. We next performed the simple dimensional reduction into the usual 10D, reproducing the BFFC system [10] for 10D, $N = 1$ supergravity, as good supporting evidence for validity of our system.

As another crucial consistency check, we have constructed a non-trivial Green-Schwarz superstring $\sigma$-model action that can couple to our 12D supergravity background. We found that the introduction of the constraint action $I_\Lambda$ guarantees the $\kappa$-invariance of the total action. Our action $I_\Lambda$ resembles the constraint action discussed in [18] bilinear in constrained fields, and the multipliers disappear from field equations. This mechanism also helps to maintain the cancellation of conformal anomalies on the string world-sheet compared with 10D superstring. We have also found the importance of the $\eta$-fermionic invariance, that together with the $\kappa$-invarinace reduces the original 32 degrees of freedom of $\theta^\mu$ into 8, in accordance with the 10D superstring theories.

The field content of our system does not produce the 11D supergravity, as seen from the absence of a forth-rank field strength. Even though we skipped in this paper, we can easily couple the supersymmetric Yang-Mills multiplet [4] to our supergravity within 12D [19], by modifying the superfield strength $G_{ABC}$ by Chern-Simons forms [20].

We stress that our theory is *not* merely a rewriting of the 10D, $N = 1$ supergravity, because extra components like $G_{ab+}$ or $T_{a+}{}^\gamma$ are non-vanishing in 12D. This is analogous to the supersymmetric Yang-Mills case in 12D [4] with the gauge and gaugino depending on the coordinate $X^+$.

The old dream about supergravity theory in 12D [7] has been now realized by the introduction of null vectors. Our investigation was motivated by the recent F-theory [1], $(2+2)$-brane [5], or S-theory [8] that suggest supergravity theories made possible by null vectors. The recent work for the globally supersymmetric Yang-Mills theory [4] also indicated the existence of supergravity with the superfield strength $G_{ABC}$. Our explicit result provides new concept that local supersymmetry can be formulated consistently with the constant null vectors. Our nice result provides strong motivation to explore other supergravity theories in 12D, in particular, formulating $N = 2$ supergravity in 12D for the field theory limit of F-theory [1] as the underlying theory for type IIB superstring. Armed with the working $N = 1$ supergravity system, the construction of the 12D $N = 2$ supergravity must be straightforward. Our result also suggests other supersymmetric theories in dimensions $\geq$ 12D, like the recently-proposed supersymmetric Yang-Mills theory in 14D [21].

Since we have established a supergravity theory in curved 12D, we can explore other possible compactifications directly from 12D into dimensions $\leq$ 10D [6]. Now with the gravitational field in curved 12D, it is easier to consider more sophisticated compactifications. It is interesting to see if any new feature arises from the extra dimensions in 12D that did not show up in the compactifications of 10D superstring theories.

The introduction of null vectors into superspace itself is not an entirely new concept. In ref. [10], we have introduced similar null vectors for constraints in superspace for $\beta$-function in the Green-Schwarz $\sigma$-model. It seems that the necessity of these null vectors in the Green-Schwarz formulation is inherent in superstring theories, which were originally formulated in the light-cone gauge. If we try to maintain the "covariance" as formally as possible in the Green-Schwarz formulation, these null vectors are to be involved, and our 12D theory is such an example.

The compatibility of our superspace backgrounds with Green-Schwarz superstring $\sigma$-model is also consistent with the prediction that heterotic or type I superstring may come from F-theory in 12D [1]. Despite of the lack of an invariant lagrangian in 12D, our Green-Schwarz formulation provides the target space effective action [22], like other similar superstring theories such as type IIB theory with no lagrangians. It is also interesting to see if the Green-Schwarz $\sigma$-model $\beta$-functions for our 12D theory provide our superfield equations [10]. Another direction is to explore a $(2+2)$-brane action [5] that may well be a more natural $p$-brane action for our 12D superspace. Studies for these directions are also under way.

We are grateful to S.J. Gates, Jr., E. Sezgin, W. Siegel, and C. Vafa for important discussions at many stages of our work presented in this paper.